\documentclass[11pt]{article}

\usepackage{cite}
\usepackage{helvet}
\usepackage{amsmath}
\usepackage{amssymb}
\usepackage{setspace}
\usepackage{graphicx}
\usepackage{empheq}
\usepackage{soul}
\usepackage{subcaption}
\usepackage{tabularx}
\usepackage{array}
\usepackage{float}
\usepackage{braket}
\usepackage[utf8]{inputenc}
\usepackage{url}
\usepackage{hyperref}

\def\be{\begin{equation}}
\def\ee{\end{equation}}

\usepackage[a4paper,top=3cm,bottom=3cm,left=2.5cm,right=2.5cm,bindingoffset=0mm]{geometry}

\begin{document}

\titlepage 

\vspace*{0.5cm}
\begin{center} 
{\Large \bf Differential Top Quark Pair Production at the LHC:\\ \vspace{3mm} Challenges for PDF Fits}\\

\vspace*{1cm}
S. Bailey, L. A. Harland--Lang \\ 
\vspace*{0.5cm}
${}^1$Rudolf Peierls Centre, Beecroft Building, Parks Road, Oxford, OX1 3PU 
\vspace*{1cm} 

\begin{abstract}

\noindent We present the results of a PDF fit to differential top quark production within the MMHT framework. We in particular consider ATLAS data in the lepton + jet and dilepton channels and CMS data in the lepton + jet channel, at 8 TeV. 
While the fit quality to the ATLAS dilepton data is good, for the CMS case we see some issues in achieving a good fit quality for certain distributions.
However, we focus on the ATLAS lepton + jet data, for which correlations of the statistical and systematic errors are provided across the four relevant distributions for PDF determination, namely $p_T^t$, $M_{tt}$, $y_t$ and $y_{tt}$. We find severe difficulties in fitting these distributions simultaneously, with particular sensitivity to the precise degree of correlation taken between the dominant two--point MC uncertainties in the data. We investigate the effect of some reasonable decorrelation of these uncertainties, finding the impact on the fit quality to be significant and the resultant gluon not negligible. This is in particular found to be larger than the effect of including NNLO QCD and NLO EW corrections in the top quark pair production cross section on the fit, motivating a closer understanding of the physics underlying these errors sources and in particular the uncertainty on the degree of correlation in them.

\end{abstract}
\end{center} 

\newpage

\section{Introduction}

The determination of proton structure via the parton distribution functions (PDFs) is an integral part of the LHC physics programme~\cite{Gao:2017yyd,Kovarik:2019xvh}. In particular, as we enter the high precision LHC era, in terms of both theory and experiment, a detailed control over all uncertainties associated with the PDFs is essential. An area of particular phenomenological relevance is the determination of the gluon PDF at high $x$, which can for example play an important role in predictions for new heavy BSM states via gluon--initiated production, but until more recently has suffered from a relative lack of direct constraints, resulting in rather large PDF uncertainties. However, with the advent of high precision LHC data at higher masses and transverse momenta this situation has in principle changed rather dramatically. As discussed in~\cite{JetFit,NNPDF,Boughezal:2017nla}, the inclusion of inclusive jet, $Z$ boson transverse momentum, and differential top quark pair production in PDF fits places important constraints on the high $x$ gluon that were lacking in earlier PDF fits. In all cases, these benefit from cutting edge high precision theory calculations at NNLO in QCD~\cite{Currie:2016bfm,Gehrmann-DeRidder:2016jns,Boughezal:2015ded,Czakon:2015owf}.

On the other hand, the inclusion of such processes in PDF fits is not without issues. In the case of jet production, it is well known~\cite{JetFit,Aad:2014vwa,decor} that the ATLAS jet production data at 7 and 8 TeV cannot be well fit across all rapidity bins. For the $Z$ boson transverse momentum distribution, issues with fitting the 7 TeV (normalized) data ~\cite{Aad:2014xaa}, and the CMS 8 TeV data~\cite{Khachatryan:2015oaa} in the highest rapidity bin have been reported~\cite{Boughezal:2017nla}, while more generally in this study an additional source of uncorrelated error, assumed to be due to residual theoretical uncertainties and possible underestimated experimental errors, has to be introduced to achieve a good fit. 

For differential top quark pair production, the first study of its inclusion in PDF fits was presented in~\cite{Guzzi:2014wia}  with approximate NNLO theory and later in~\cite{NNPDF}, where the full NNLO calculation was used in the context of a global fit with rather encouraging results. Here, 8 TeV data from ATLAS~\cite{ATLAS_lep_jet}, and CMS~\cite{CMS_lep_jet}, both in the lepton + jet channel, were considered. These data were presented differentially in the top quark pair invariant mass, $m_{t\overline{t}}$, and rapidity, $y_{t\overline{t}}$, and the individual top quark/antiquark transverse momentum, $p_\perp^t$, and rapidity, $y_t$. However, as the corresponding statistical correlations across these different distributions were not then available, only one of these distributions from each dataset could be included at once. In all of the above cases, the data were corrected back to the top quark parton level.

This situation has changed quite recently, with the ATLAS collaboration providing the corresponding statistical correlations for their 8 TeV data~\cite{ATL-PHYS-PUB-2018-017}, while the systematic errors were provided in the original data. This therefore allows a simultaneous fit to all four distributions to be performed for the first time, in principle providing a greater discriminating power in the determination of the high $x$ gluon. However, as discussed in~\cite{ATL-PHYS-PUB-2018-017}, when even two distribution are fit simultaneously within the context of the ATLAS PDF fit, only a very poor fit quality can be achieved, while a similar observation has been made by the MMHT~\cite{LHLtalkATLAS,Thorne:2019mpt} and CT~\cite{Hou:2019gfw} collaborations. This therefore casts doubt on the reliability of including such data in PDF fits. It should in particular be emphasised that this issue is not bypassed by fitting to one individual distribution alone, which would artificially mask the issues in fitting the complete dataset, while also risking introducing a bias into the fit, through the particular choice of distribution that is made.

Further study of the above effects is clearly essential, and indeed in~\cite{ATL-PHYS-PUB-2018-017}, the source of this poor fit quality was traced to a small subset of dominant two--point systematic uncertainties associated with the choice of Monte Carlo generator or input parameters, used in the data unfolding to the parton--level results which are included in the PDF fit. A rather significant improvement in the fit quality, with a relative stability in the resultant gluon, was then observed by introducing some decorrelation in a given systematic error between distributions, reminiscent of the studies\cite{JetFit,decor} for the case of jet production. On the other hand, this study was only performed within the ATLAS PDF framework, fitting to a limited dataset and with a rather restrictive parametrisation, while only fits to only two distributions were made rather than the complete set of four. 

In this paper, we therefore present the first consistent fit to the full ATLAS dataset, within the global MMHT framework. We study in detail the variation in the fit quality that comes from fitting individual datasets in comparison to the combination, the effect that introducing some degree of decorrelation in the two--point experimental systematics has, and the impact of this on the resultant gluon PDF. We consider the impact of including the NNLO QCD and NLO EW corrections to the top quark production matrix elements on the gluon, finding that these are rather smaller than those due to the prescription for treating the experimental systematic correlations. As we will discuss, this is equally true if an individual distribution alone is fit to, which in effect implicitly assumes a complete decorrelation of all systematics within distributions. 

We in addition consider the ATLAS data in the dilepton channel, finding that here a good fit to the individual distributions can be achieved. While a fit to the total dataset can not be performed, as the corresponding statistical correlations are not available, this suggests a more reliable fit might be achieved here. Finally, we consider the CMS lepton + jet data, taking the rapidity distributions, $y_t$ and $y_{tt}$, for concreteness. We find that a fair description can be achieved in the latter case, while the fit quality is rather poor in the former.

The outline of this paper is as follows. In Section~\ref{sec:datasets} we describe the datasets we include in the fit. In Section~\ref{sec:ATLAS} we describe the fit to the ATLAS lepton + jet data, including a detailed studied of the impact of the decorrelation of systematic errors and the impact on the gluon PDF. In Section~\ref{sec:dilepton} we describe the fit to the ATLAS dilepton data. In Section~\ref{sec:CMS} we describe the fit to the CMS lepton + jet data. In Section~\ref{sec:comb} we describe a combined fit to all three datasets. Finally, in Section~\ref{sec:conc} we conclude.

\section{Data Sets and Theoretical Calculation}\label{sec:datasets}

We consider ATLAS 8 TeV data on differential top quark production in lepton + jet~\cite{ATLAS_lep_jet} and dilepton~\cite{ATLAS_dilep} channels. The lepton + jet data are presented as single differential distributions with respect to a number of parton--level kinematic variables. The four which are useful for this analysis are: the average transverse momentum of the $t(\bar{t})$-quark ($p_T$), the invariant mass of the $t\bar{t}$ pair ($M_{tt}$), the average rapidity of the $t(\bar{t})$-quark ($y_t$) and the rapidity of the $t\bar{t}$ pair ($y_{tt}$). Correlations between individual systematic errors are available both within and between distributions, as well as a full statistical correlation matrix. For the dilepton dataset, this is again presented as single differential distributions with respect to two relevant kinematic variables for PDF fitting, namely the rapidity and invariant mass of the $t\bar{t}$ pair, however the systematic and statistical correlations are not available between these distributions. In this study we will consider individual fits to the two cases.

We in addition consider the CMS 8 TeV in the lepton + jets channel~\cite{CMS_lep_jet}, which are given in terms of the same kinematic variables as in the ATLAS case. These data are presented as normalized distributions, which we simply multiply by the total cross section measured in the same channel~\cite{cms_total}, taking the statistical, systematic and luminosity errors associated with this as fully correlated between all bins. The statistical correlation matrix within each distribution is provided, but not across distributions. Therefore, as in the ATLAS dilepton case, only a single distribution at a time can be fit.

For the theoretical input, predictions at NNLO in QCD for the ATLAS lepton + jets and CMS data are provided in~\cite{Czakon:2017dip} as fastNLO grids \cite{Kluge:2006xs}, while EW K--factors are evaluated using the predictions of~\cite{NNLOxEW}, as provided using the PDF4LHC15LUXqed~\cite{Manohar:2016nzj} set. As the K--factors should be relatively insensitive to the PDF set used, this should provide a good estimate for the current case. For the ATLAS dilepton data, NLO QCD Applgrids~\cite{Carli:2010rw}, NNLO QCD K--factors and EW NLO K--factors were supplied by the ATLAS collaboration~\cite{k_facs}. As NNLO grid files for the CMS dilepton data are not currently available, we do not consider these in what follows.

\section{Lepton + Jets Channel at ATLAS}\label{sec:ATLAS}

We first consider the ATLAS data at 8 TeV, in the lepton + jet channel, using the recently provided full statistical correlation matrix~\cite{ATL-PHYS-PUB-2018-017}. As discussed in the introduction, significant issues in achieving a good fit to these data have been found by ATLAS, with similar observation made by CT and MMHT. We analyse this in detail below.

\subsection{Fitting within the MMHT framework}

As a baseline, we use the MMHT14 PDF set~\cite{mmht}, but including the final HERA I + II combination data~\cite{HERA} along with some new total $t\bar{t}$ cross section data, and finally including some minor improvements to the fitting code. When we add in the ATLAS data to this, we remove the corresponding total cross section measurement to avoid double counting, though the effect of this is rather small.

In the fit we use the figure of merit:
\begin{equation}
\chi^2 = \sum_{i,j=1}^{N_{\text{data}}} \left( D_i-T_i-\sum_{\alpha = 1}^{N_\text{sys}} \beta_{i,\alpha} \lambda_\alpha \right) \left( \text{cov}_{\text{stat}} \right)_{ij}^{-1} \left( D_j-T_j-\sum_{\gamma = 1}^{N_\text{sys}} \beta_{j,\gamma} \lambda_\gamma \right) + \sum_{\alpha = 1}^{N_{\text{sys}}} \lambda_\alpha^2,
\label{chi2 eq}
\end{equation}
where $D_i$ and $T_i$ are the data and theory predictions, $\beta_{i,\alpha}$ are the systematic errors, $\text{cov}_{\text{stat}}$ is the statistical covariance matrix and $\lambda_{\alpha}$ are the nuisance parameters, due to the experimental correlated systematic errors. Here we have symmetrized the systematic errors and have shifted the central values of the data points accordingly to account for this. While in principle one can minimise the above $\chi^2$ analytically, by profiling with respect to these nuisance parameters, we have kept them explicit here to aid the discussion which follows.

The $\chi^2$ per data point after fitting, for the combined fit to all four distributions, as well as various individual fits to be discussed further below, is shown in Table~\ref{chi2 1}. We can see immediately that the fit quality in the combined case is extremely poor, with a $\chi^2$ of $\sim$ 7 per point. If we consider instead a subset of two distributions, as in the ATLAS study~\cite{ATL-PHYS-PUB-2018-017}, the fit quality is better but still very poor. To be concrete, we show the result of a fit to the $p_T$ and $M_{tt}$ distributions, though for other choices the results are rather similar. We also show results excluding the statistical correlations, to asses their impact, and find that these can play an important role, but that even excluding them artificially, the fit quality is still poor. Finally, we find that even fitting to the distributions individually, while a good description of the $p_T$ and $M_{tt}$ cases is possible, the fit to the rapidity variables is still rather poor, with a $\chi^2$ $\sim 3$ per point. This is consistent with the results of the ATLAS study~\cite{ATL-PHYS-PUB-2018-017}, as well as CT~\cite{Hou:2019gfw}, but interestingly not with the NNPDF study of~\cite{NNPDF}, where a good description of all individual distributions was found. Here, the impact of statistical correlations within the distributions, which were not available for the NNPDF study, is seen to be minimal, and so cannot explain this difference.

\begin{figure}
\centering
\begin{subfigure}{0.45\textwidth}
\includegraphics[width=\textwidth]{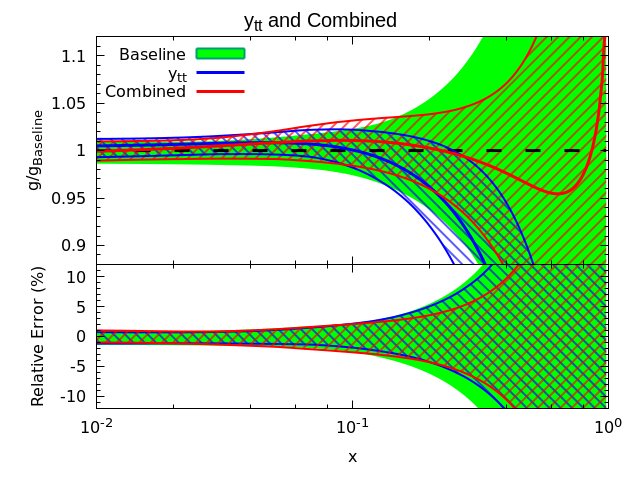}
\end{subfigure}
~ 
\begin{subfigure}{0.45\textwidth}
\includegraphics[width=\textwidth]{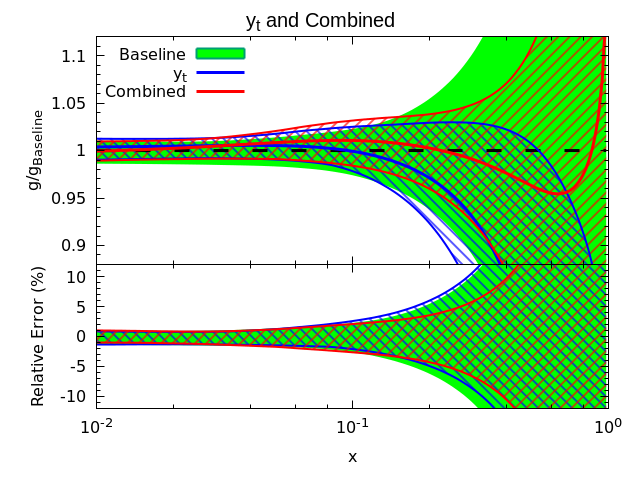}
\end{subfigure}
\begin{subfigure}{0.45\textwidth}
\includegraphics[width=\textwidth]{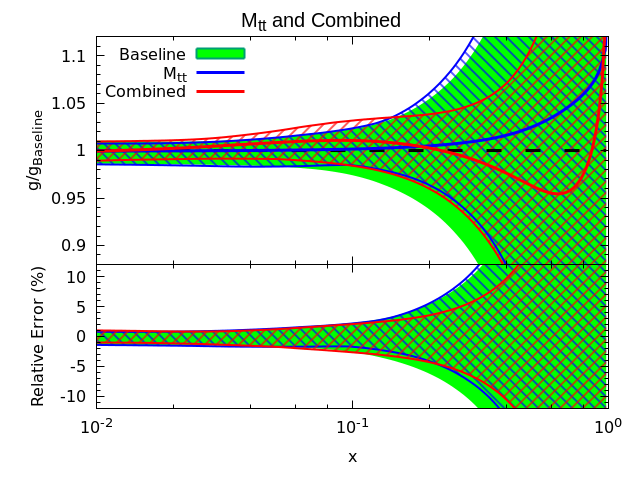}
\end{subfigure}
~
\begin{subfigure}{0.45\textwidth}
\includegraphics[width=\textwidth]{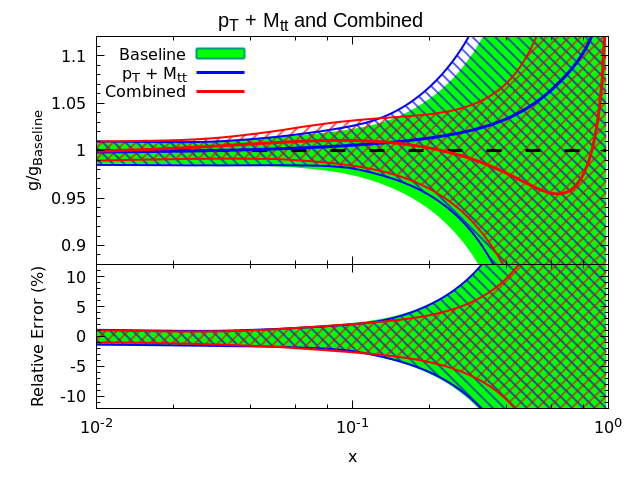}
\end{subfigure}
\caption{Impact of the ATLAS 8 TeV lepton + jet data on the gluon PDF. Results for fits to individual distributions as well as the combined $p_T+M_{tt}$ case are shown, while the result from a combined fit to all four distributions is shown for comparison in all cases.}
\label{pdf 1}
\centering
\end{figure}

We now consider the impact on the gluon itself, shown in Fig.~\ref{pdf 1} for fits to different distributions, both individual and in combination. Broadly speaking, we can see that as expected the data has a noticeable impact on the high $x$ gluon, both in terms of the central value and uncertainty. However, on closer inspection we can see that the impact of the individual fits on the gluon is somewhat different, with the rapidity variables tending to decrease the gluon at high $x$, while the $p_T$ and $M_{tt}$ tend to pull in the opposite direction. While the resulting gluon PDFs in all cases agree within their error bands, indicating that there is no strong tension between them, nonetheless a difference in the overall pulls is clear. This difference in trend is also observed in the ATLAS study~\cite{ATL-PHYS-PUB-2018-017}. 

\begin {table}
\begin{center}
\begin{tabular}{|c|c|c|} 
\hline
Distribution& Statistics Correlated & Statistics Uncorrelated \\
\hline
$p_T$ & 0.53 & 0.50 \\
\hline
$y_t$ & 3.12 & 3.16 \\
\hline
$y_{tt}$ & 3.51 & 3.51 \\
\hline
$M_{tt}$ & 0.70 & 0.60 \\
\hline
$p_T+M_{tt}$ & 5.73 & 2.47 \\
\hline
Combined & 7.00 & 3.28 \\
\hline
\end{tabular}
\caption{$\chi^2/N_{\rm data}$ values for fits to different distributions within the ATLAS 8 TeV lepton + jet data, as well as for the combined fit to all four distributions. The left (right) columns correspond to the case that the statistical correlations are included (excluded).}
\label{chi2 1} 
\end{center}
\end{table}

\subsection{Understanding the poor fit quality: the role of correlated systematic errors}\label{decor sec}
In the previous section we observed a very poor fit quality to the ATLAS 8 TeV lepton + jet data when two or more distributions were fit simultaneously. Here, we wish to try and understand this effect further. In particular, as discussed in the ATLAS analysis~\cite{ATL-PHYS-PUB-2018-017}, the experimental uncertainties for this dataset are completely dominated by the experimental systematic errors. For the rapidity distribution the three largest sources of systematic error are $\sim 3 -6 $ \% (hard--scattering model), $\sim 6-9$ \% (ISR/FSR) and $\sim 3$ \% (parton shower model), depending on the distribution and rapidity bin, while the statistical uncertainty is only $\sim 0.6 -1.3$ \%. We will return to these systematic errors shortly, but for now simply note that the fit quality will naturally be driven by the precise treatment of the correlations in these errors. In particular, as we will see it is the correlation across different distributions that drives the observed deterioration in fit quality. 

\begin{figure}
\centering
\begin{subfigure}{0.5\textwidth}
\includegraphics[width=\textwidth]{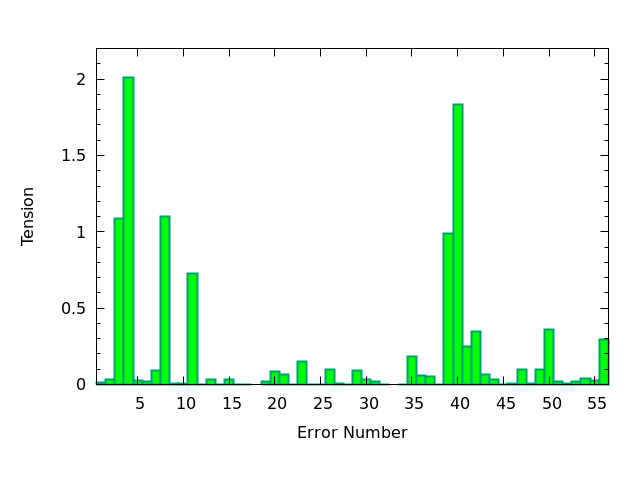}
\end{subfigure}
\caption{The tension between shifts for the ATLAS 8 TeV lepton + jet data, between the individual distributions. This is calculated as $(1/6)\sum_{j=1}^{4}\sum_{l>j}^{4}(\lambda_{i,j}-\lambda_{i,l})$ where $\lambda_{i,j}$ is the $i^\text{th}$ nuisance parameter calculated when fitting the $j^\text{th}$ distribution.}
\label{tension}
\centering
\end{figure}

To investigate the above effect further, we first follow the study of~\cite{JetFit} and evaluate the variation in the preferred values of the corresponding systematic shifts in \eqref{chi2 eq} when fitting to the different distributions individually. Any significant tension between these values will then indicate that when fitting the distributions in combination, some deterioration in the fit quality will occur, and moreover that a less restrictive degree of correlation between the distribution for the corresponding shift may be preferred by the data/theory comparison. However it should be stressed that this can clearly only be taken as a guide, as one must of course also consider whether it is reasonable to consider a different correlation scenario for the corresponding shift, or whether this is already precisely determined by the experimental analysis, with no further decorrelation being possible. The result is shown in Fig.~\ref{tension}, and we find that the jet energy scale (4) and ISR/FSR (40) errors provide the largest source of tension. As the correlation in the former may reasonably be assumed to be well determined experimentally, we will not consider this further. On the other hand, in the latter case this is less clear, as we shall now discuss (see also~\cite{ATL-PHYS-PUB-2018-017}).

In fact, further investigation reveals that the errors relating to the hard-scattering model and the parton shower are very similar to the ISR/FSR error for most bins, as well as being the largest systematic errors, as mentioned above. This suggests that there is some redundancy in our analysis, with the large tension in one of these errors masking the tension in the others. Thus is reasonable to consider all three of these errors as a potential source of tension. Now, the important point here is that in all three cases these are 2--point errors evaluated using two choices of Monte Carlo (MC) generator or generator inputs. In particular, a given choice of MC generator and input parameters must be taken to unfold the observed data back to the parton--level distributions which we use for our PDF fits. To evaluate the systematic uncertainty due to this choice, a second generator/set of input parameters is used to unfold MC signal events generated with the default MC. The difference between this result and parton--level signal is then assigned to be the systematic uncertainty in each bin. More specifically, in the case of the hard scattering the default signal is generated with \texttt{MC@NLO} + \texttt{Herwig} and the unfolding is performed with \texttt{POWHEG}+\texttt{Herwig}, while for the parton shower the difference is between \texttt{POWHEG}+\texttt{Herwig} and \texttt{POWHEG}+\texttt{Pythia}, and for ISR/FSR a variation in the input parameters of \texttt{POWHEG}+\texttt{Pythia} is taken. 

Now, crucially in all cases the correlation in the systematic error is also taken from this procedure, with it being assumed that any correction factor evaluated by this two--point procedure should be applied in a fully correlated way across all bins. To some extent this is a reasonable assumption: if a different choice of MC generator/input parameters for the unfolding leads to a larger parton--level result at larger $m_{t\overline{t}}$ and $p_t$, say, then there may well be physics or kinematic reasons for these effects to be correlated. However, there is clearly some uncertainty in this degree of correlation, and indeed if one performed the same study as above with a third choice of MC/input parameters, there is no reason to assume that in all cases the corresponding correction would always lie within the correlation two--point band. More specifically, when determining the systematic errors, it is assumed that the result of a potential third MC or choice of input parameters would be perfectly describable by taking some particular value of the single $\lambda_\alpha$ shift assigned to this error source. Put another way, the effect of this is constrained to be a linear combination of the two baselines used to evaluate the systematic error, with the combination being the same across all bins. This may not necessarily be the case, and indeed given the issues with fit quality observed above, which as we will see are to a large extent driven by these correlated systematics, it is reasonable to investigate the effect of loosening the assumed correlation scenarios for these.

\begin{figure}
\centering
\begin{subfigure}{0.5\textwidth}
\includegraphics[width=\textwidth]{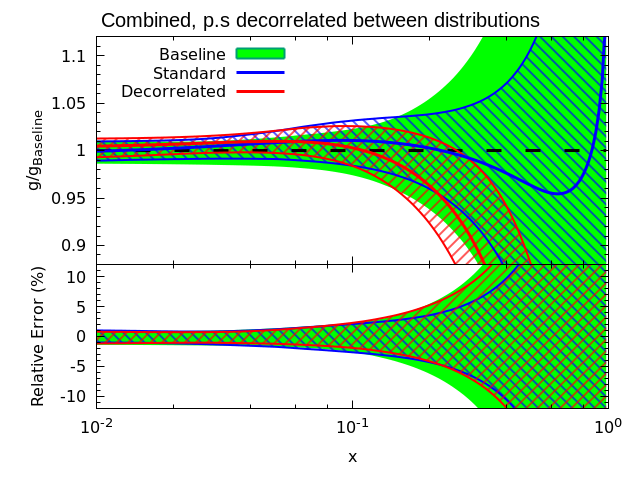}
\end{subfigure}
\caption{Impact of the ATLAS 8 TeV lepton + jet data on the gluon PDF. The result of the default to the combined dataset is shown, as well as decorrelating the parton shower systematic uncertainty between the four distributions.}
\label{decor sets pdf}
\centering
\end{figure}

\begin {table}
\begin{center}
\begin{tabular}{|c|c|c|} 
\hline 
Distribution& p.s. correlated & p.s. decorrelated \\
\hline
Combined & 7.00 & 1.80 \\
\hline
$p_\perp^t + M_{tt}$ & 5.73 & 0.66 \\
\hline
\end{tabular}
\caption{$\chi^2/N_{\rm data}$ values for fits to the ATLAS 8 TeV lepton + jet data, including decorrelation of the parton shower systematic uncertainty described in the text.}
\label{Table:decor} 
\end{center}
\end{table}

A further related issue is the assumption implied in \eqref{chi2 eq} that the corresponding systematic uncertainties are standard symmetric Gaussian sources of error. As the discussion above makes clear, these two--point systematic uncertainties, as well as not providing a well defined degree of correlation, are in addition expected to be inherently non--Gaussian. For example, there is no particular reason to associate a 1--$\sigma$ uncertainty, as opposed to a smaller or larger allowed variation, with the uncertainty bands calculated in the above way. These error sources can in addition be rather asymmetric. While the parton shower and hard scattering errors are provided by ATLAS after symmetrisation, the ISR/FSR error does have some degree of asymmetry, particularly in the high invariant mass bins (though less so in the for example the rapidity distributions, where already we see a poor individual fit quality). In the latter case we choose to simply symmetrize, as described above, but this is only one possible choice. More generally, it would be desirable to investigate the impact of different assumptions about the above effects, and of including a non--Gaussian probability distribution for these error sources. From a practical point of view this makes a direct application of \eqref{chi2 eq} much less tractable, as a simple analytic minimisation with respect to the corresponding nuisance parameters is not possible. A full analysis of these issues is beyond the scope of the current study, but is clearly important in the future.

\begin{figure}
\centering
\begin{subfigure}{0.5\textwidth}
\includegraphics[width=\textwidth]{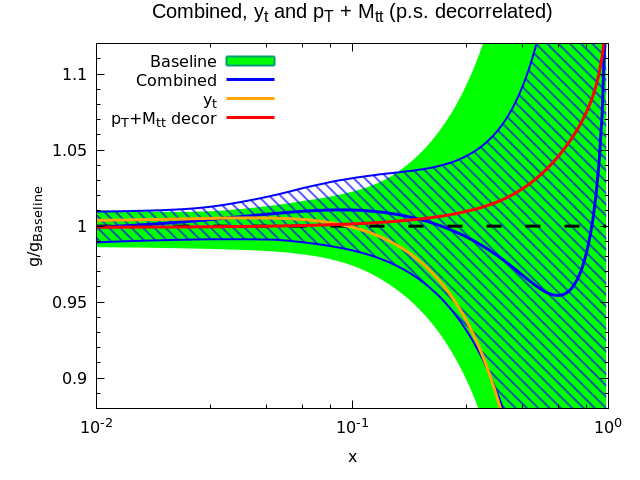}
\end{subfigure} 
\caption{As in Fig.~\ref{pdf 1}, for fits to different distributions taken in the literature: $y_t$ and $p_T+M_{tt}$ (with decorrelation of the parton shower systematic error), as well as the combined fit.}
\label{pdf 2}
\centering
\end{figure}

Here, we therefore concentrate solely on the question of correlation, and in particular investigate the effect of loosening the correlation in the systematic error associated with parton shower, guided by the discussion above, and following the choice made in the ATLAS analysis~\cite{ATL-PHYS-PUB-2018-017}. We in particular consider the same combined fit to the four distributions as before, including all statistical correlations, but now completely decorrelating the parton shower systematic error within the four distributions. That is, we split this source of uncertainty into four sources, which each only contributing in a given individual distribution, while keeping all correlations within the distributions for now untouched. The result is shown in Table~\ref{Table:decor}, and we can see that the effect is dramatic, leading to a factor of $\sim 4$ decrease in the $\chi^2$. This confirms the discussion above, showing that indeed the assumed degree of correlation in these two--point systematics has a significant impact on the fit quality. In Fig.~\ref{decor sets pdf} we show the impact this has on the corresponding gluon. We can see that there is some non--negligible difference in the result, such that the gluon is not completely stable under different treatments of the systematic correlations. Moreover, we can see that now the pull on the gluon is in fact close to the pull from the individual fits to the rapidity variables, see Fig.~\ref{pdf 1}, suggesting these distributions are having a more dominant effect after the decorrelation. 

We summarise the situation, directly comparing to choices that have been made in the literature, in Fig.~\ref{pdf 2}. Here we show the impact of fits to difference distributions, namely the individual rapidity $y_t$, the combined $p_T+M_{tt}$ (including parton shower decorrelation), and the fit to all four distributions, on the same plot. A rather large spread in results is observed, emphasising the need for further investigation.

\begin{figure}
\centering
\begin{subfigure}{0.5\textwidth}
\includegraphics[width=\textwidth]{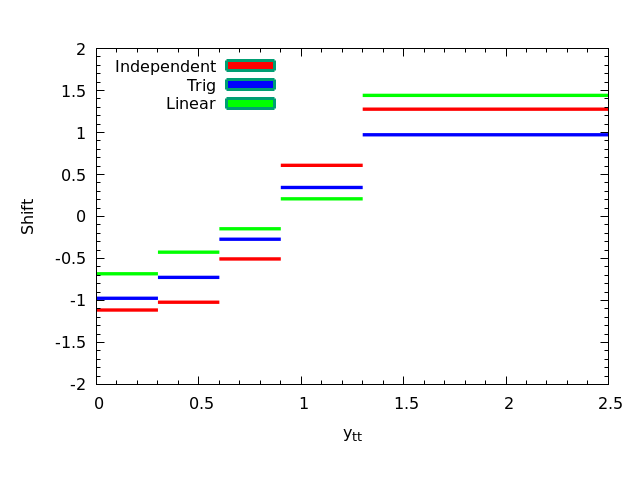}
\end{subfigure}
\caption{Values for individual shift parameters ($\lambda_i$ in Eq.~\eqref{chi2 1}) for the parton shower error when fitting the $y_{tt}$ distribution of the ATLAS 8 TeV lepton + jet data, while decorrelating the error between each data point.}
\label{shifts}
\centering
\end{figure}

Given that the rapidity distributions appear to be driving the gluon impact after decorrelating across distributions, it is worth investigating the fit these distributions individually further. In particular, given that the individual fit quality to these is rather poor, see Table~\ref{chi2 1}, we should investigate the impact of the above correlations within the rapidity distributions on the extracted gluon. According to the arguments above, it is equally possible that the degree of correlation taken within the rapidity distribution, as well as across distributions, for the two--point MC uncertainties may be overly restrictive. To investigate this, we will consider the $y_{t\bar{t}}$ distribution, though similar results are found for $y_{t}$. We first allow the parton--shower uncertainty to be completely uncorrelated across the bins of the rapidity distribution, with one corresponding nuisance parameter per bin, that are each allowed to shift independently. The result is shown in Fig.~\ref{shifts}, and we can see that rather a smooth shape in the shifts is preferred. While in all cases the magnitude of the shift is $\sim 1$, indicating that the shifts are not being pushed particularly outside the $1\sigma$ bands corresponding to the magnitude of the systematic errors in each bin, the shape is clearly quite different from that required by the default correlation prescription.

Motivated by this, we allow for a degree of decorrelation within the rapidity distribution, following the approach of~\cite{decor}, where a similar decorrelation was introduced for various sources of experimental systematic and theoretical uncertainty within the rapidity and $p_\perp$ bins of the ATLAS 8 TeV inclusive jet data. In particular, we split the parton shower error into two components for each bin, with the first error following some function of the distribution variable, and the other chosen such that they add in quadrature to the original error. We choose two simple approaches, a linear function with the errors given by:
\begin{align}
\beta_i^{(1)} &= \left( \frac{y_{tt,i}-y_{tt,\text{min}}}{y_{tt,\text{max}}-y_{tt,\text{min}}}\right) \beta_i^{\text{tot}}, & \beta_i^{(2)} &= \left[ 1 - \left( \frac{y_{tt,i}-y_{tt,\text{min}}}{y_{tt,\text{max}}-y_{tt,\text{min}}}\right)^2 \right]^{\frac{1}{2}} \beta_i^{\text{tot}}
\end{align}
and a trigonometric function given by:
\begin{align}
\beta_i^{(1)} &= \cos \left[ \pi \left( \frac{y_{tt,i}-y_{tt,\text{min}}}{y_{tt,\text{max}}-y_{tt,\text{min}}}\right)\right] \beta_i^{\text{tot}} , & \beta_i^{(2)} &= \sin \left[ \pi\left( \frac{y_{tt,i}-y_{tt,\text{min}}}{y_{tt,\text{max}}-y_{tt,\text{min}}}\right)\right] \beta_i^{\text{tot}}.
\end{align}

We note that the trigonometric decorrelation has the advantage of naturally adding in quadrature to the magnitude of the original error, while allowing some additional freedom in the relative sign of the shift across the distribution. Although it may seem preferable to instead use a factor of $\pi/2$ in the arguments for sine and cosine to preserve symmetry, the above choice provides a better description of the desired shifts and a higher quality fit.

\begin{figure}
\centering
\begin{subfigure}{0.45\textwidth}
\includegraphics[width=\textwidth]{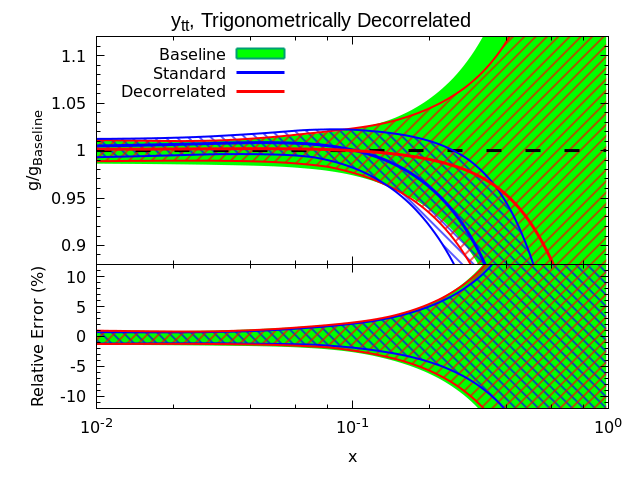}
\caption{}
\label{decor_ytt}
\end{subfigure}
~
\begin{subfigure}{0.45\textwidth}
\includegraphics[width=\textwidth]{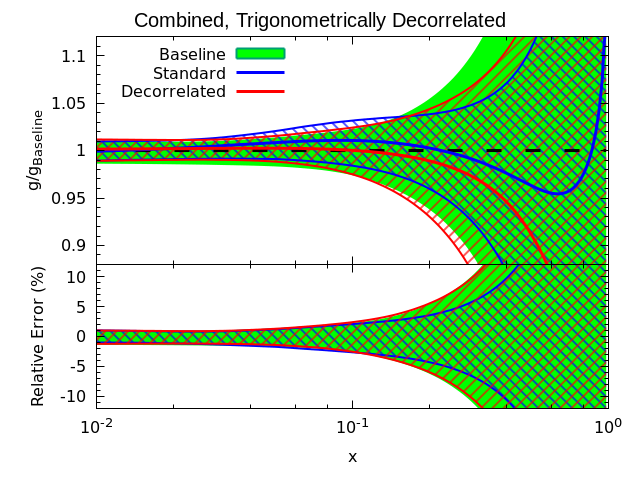}
\caption{}
\label{decor_all_ytt}
\end{subfigure}
\caption{Fits to the $y_{tt}$ distribution and all distributions together along with different decorrelation procedures along $y_{tt}$, compared to fits with no decorrelation.}
\label{ytt decor pdfs}
\centering
\end{figure}

The results of the fit to the $y_{tt}$ distribution are shown in Table~\ref{Table:decor}. We can see that the effect of the trigonometric decorrelation is significant, reducing the $\chi^2$ per point to $\sim 1$, over a factor of 3 lower than in the default case. For the linear case the effect is rather less pronounced, though some reduction is seen, suggesting that the trigonometric decorrelation allows for the data to be fit more consistently within our framework. This is demonstrated in Fig.~\ref{shifts}, where it can be seen that this decorrelation more closely represents the shifts seen in the completely decorrelated case. In Fig.~\ref{decor_ytt} we show the impact on the gluon PDF for the trigonometric case. We can see that some shift in the central value of the gluon occurs, albeit within PDF uncertainties. More significantly, the resultant gluon uncertainty is rather larger, and indeed very similar to the uncertainty on the baseline set. It would therefore appear that by allowing for a weaker degree of correlation, much of the constraining power of this particular distribution that appeared to be present in the default case (albeit with a rather poor fit quality, raising questions about the reliability of this) has been washed out. In essence, by allowing a greater freedom in the correlation of the systematic error associated with parton shower, to a large extent it appears that the data can be fit simply by shifts of the corresponding nuisance parameters, resulting in the end in a rather small constraining power on the gluon itself. We note that for this effect to occur it is not necessary for the original fit quality to be bad; even if the original fit quality is good, nonetheless if the correlations taken for such dominant sources of systematic error are too constraining this may still result in an artificially large constraint on the corresponding gluon PDF.

\begin {table}
\begin{center}
\begin{tabular}{|c|c|c|} 
\hline
Distribution& p.s. correlated & p.s. decorrelated \\
\hline 
$y_{t\overline{t}}$ (linear) & 3.51 & 2.62 \\
\hline
$y_{t\overline{t}}$ (trig.) & 3.51 & 1.02 \\
\hline
Combined ($y_{t\overline{t}}$ (trig.)) & 7.00 & 1.62 \\
\hline
\end{tabular}
\caption{$\chi^2/N_{\rm data}$ values for fits to the ATLAS 8 TeV lepton + jet rapidity distributions, including decorrelation of the parton shower systematic uncertainty described in the text.}
\label{Table:decory} 
\end{center}
\end{table}

Finally, we should consider the impact of this within a fit to the combined distributions, that is by decorrelating the parton shower error between each set and along the $y_{tt}$ distribution using the trigonometric decorrelation. The result, shown in Table~\ref{Table:decory}, is a $\chi^2$ of 1.62 per point, only marginally better than correlating solely between sets. The impact on the gluon is shown in Fig.~\ref{decor_ytt} (b), where we see a relatively small shift in the central value, but again an increase in the corresponding PDF uncertainty to the level of the baseline set. Thus again in this case the constraining power of the data appears to be somewhat washed out. We note that the fit quality of 1.62 per point is still relatively poor, being driven by the fact that the $y_t$ distribution is also not well described, see Table~\ref{chi2 1}. We have investigated the effect of introducing the same decorrelation as above to the $y_t$ distribution, and indeed find an improvement in the fit, while the gluon is quite similar to Fig.~\ref{decor_ytt}. Simply omitting this rapidity distribution gives a similar result to this.

\subsection{Theoretical precision}

In the previous section, we have observed that the degree of correlation taken for the dominant two--point systematic errors (taking the parton shower as a test case) can have a rather large effect on the extracted gluon from the fit to the ATLAS lepton + jet data. This is clearly a cause of concern, in particular in light of the high precision being aimed for in the PDF fit. In particular, as discussed in the introduction, we can perform a fit using cutting edge NNLO in QCD theoretical predictions for the parton--level top quark pair production process~\cite{Czakon:2015owf}, and including NLO EW corrections on top of this. The theoretical precision in these predictions is therefore high, but while one would be tempted to claim a similar degree of precision in the corresponding PDF extraction, the above results cast some doubt on this, at least for the ATLAS lepton + jet data.

\begin{figure}
\centering
\begin{subfigure}{0.45\textwidth}
\includegraphics[width=\textwidth]{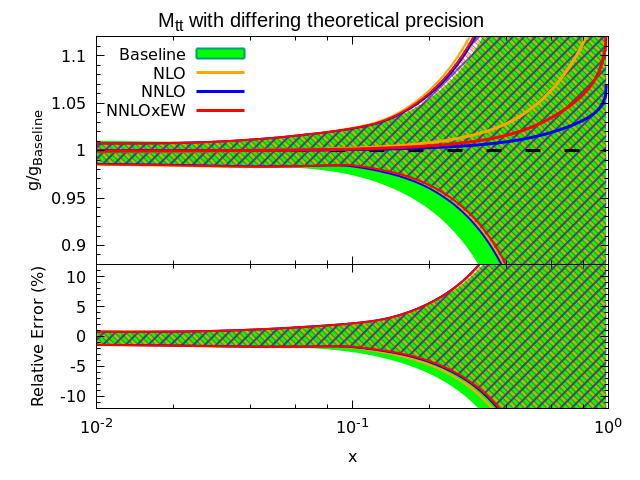}
\caption{}
\label{mtt_theory}
\end{subfigure}
~
\begin{subfigure}{0.45\textwidth}
\includegraphics[width=\textwidth]{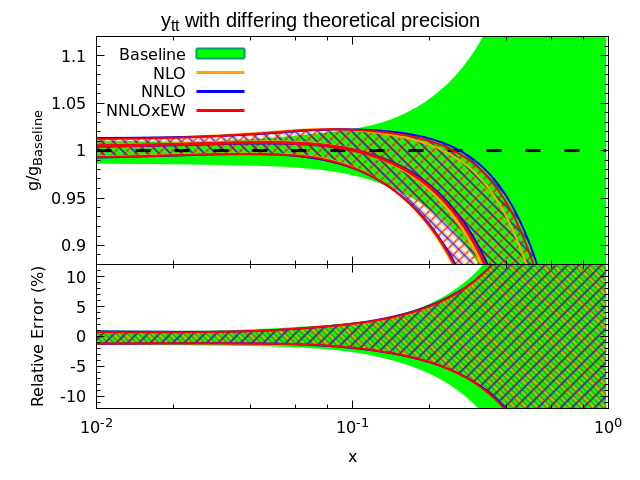}
\caption{}
\label{ytt_theory}
\end{subfigure}
\caption{Impact of the ATLAS 8 TeV lepton + jet data on the gluon PDF, for different levels of precision in the theoretical prediction for the top quark pair production cross section, when fitting the $M_{tt}$ (left) and $y_{tt}$ (right) distributions.}
\label{theory pdfs}
\centering
\end{figure}

To investigate this point further, we evaluate the impact of including NNLO QCD and NLO EW corrections to the top quark pair production matrix element on the extracted gluon PDF. To be specific, we will evaluate the top quark pair production cross section at NLO, NNLO and NNLO$\times$EW order, while keeping all other fit settings the same, i.e. we in particular use NNLO PDFs throughout, in order to isolate the effect of these corrections. The impact on the fit quality is shown in Table~\ref{tab:chiorder}, and one can see that the NNLO QCD corrections lead to a better description of the data in comparison to the NLO, as we would hope for, while interestingly the EW corrections leads to some deterioration in the fit quality. On the other hand, the impact of these effects on the extracted gluon PDF when fitting to the $m_{tt}$ and $y_{tt}$ distributions is shown in Figs.~\ref{mtt_theory} and 
~\ref{ytt_theory}, and is seen to be rather small; for the other distributions the difference is smaller still. 

\begin{figure}
\centering
\begin{subfigure}[b]{0.5\textwidth}
\includegraphics[width=\textwidth]{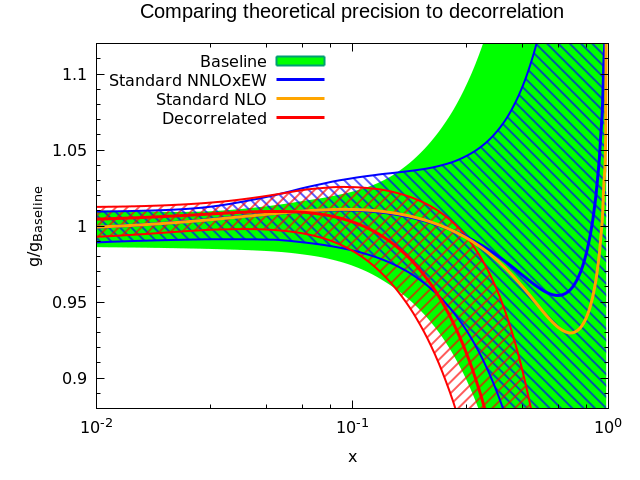}
\end{subfigure}
\caption{Extracted gluon from a fit to the ATLAS 8 TeV lepton + jet data (all four distributions), from a fit including NNLO$\times$EW in the cross section with (`decorrelated') and without (`standard') decorrelation of the parton shower systematic error, and from a fit with pure NLO in the cross section calculation, without decorrelation.}
\label{theory decor pdfs}
\centering
\end{figure}

Further, in Fig.~\ref{theory decor pdfs}, we compare the impact on the gluon of the theoretical precision in the cross section calculation to the treatment of the parton shower error in the data, for the case that all four distributions are fit simultaneously. We in particular show the result of a fit including NNLO$\times$EW corrections, with and without decorrelation, as well as the result of a fit including NLO corrections alone, and with no decorrelation. We can see that the difference between the NLO and NNLO$\times$EW cases is indeed rather smaller than the difference between the NNLO$\times$EW case with decorrelation to the case without. Thus if we are to extract the gluon PDF from such data with an accuracy to match the high precision provided by the NNLO$\times$EW calculation of the underlying cross section, it will be crucial to have a clearer understanding of these dominant two--point systematic uncertainties and the uncertainty on their correlation.

\begin {table}
\begin{center}
\begin{tabular}{|c|c|c|c|} 
\hline
Distribution & NLO & NNLO & NNLO+EW \\
\hline
$p_T$ & 0.65 & 0.36 & 0.53\\
\hline
$y_t$ & 2.99 & 2.98 & 3.12 \\
\hline
$y_{tt}$ & 4.06 & 3.30 & 3.51 \\
\hline
$M_{tt}$ & 1.33 & 0.57 & 0.70\\
\hline
All & 7.88 & 6.61 & 7.00 \\
\hline
\end{tabular}
\caption{$\chi^2/N_{\rm data}$ values for fits to different distributions within the ATLAS 8 TeV lepton + jet data, using NLO, NNLO and NNLO+EW theory for the top quark pair production cross section.}
\label{tab:chiorder}
\end{center}
\end{table}

\section{Dilepton Channel at ATLAS}\label{sec:dilepton}

\begin {table}
\begin{center}
\begin{tabular}{|c|c|} 
\hline
Distribution & $\chi^2/N_{\rm data}$ \\
\hline
$M_{tt}$ & 0.06 \\
\hline
$y_{tt}$ & 0.66 \\
\hline
\end{tabular}
\caption{$\chi^2/N_{\rm data}$ values for fits to the ATLAS 8 TeV dilepton data.}
\label{Table:dilep} 
\end{center}
\end{table}

\begin{figure}
\centering
\begin{subfigure}{0.5\textwidth}
\includegraphics[width=\textwidth]{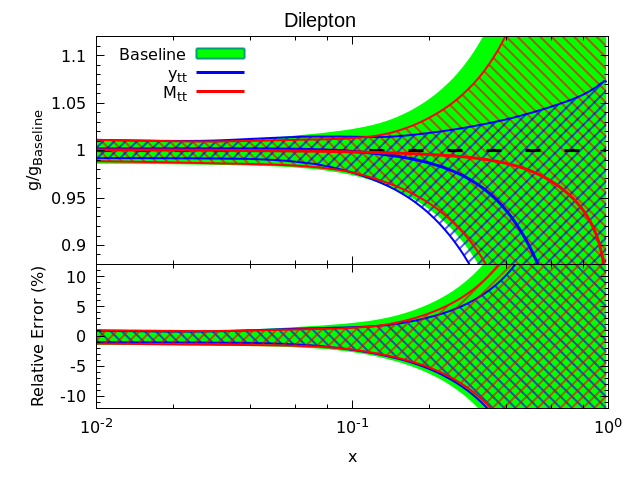}
\end{subfigure}
\caption{Impact of the ATLAS 8 TeV dilepton data on the gluon PDF, for the $y_{tt}$ and $M_{tt}$ distributions.}
\label{dilep}
\centering
\end{figure}

In the preceding sections, we have considered a fit the ATLAS 8 TeV data collected in the lepton + jet channel, however data in the dilepton channel~\cite{ATLAS_dilep} at 7 and 8 TeV are also available, of which we fit to the higher precision latter dataset. Unfortunately, the statistical and systematic correlations are not provided across distributions, and hence a complete study cannot be performed. We therefore instead consider fit to the $y_{tt}$ and $M_{tt}$ distributions individually. The results are shown in Table~\ref{Table:dilep}, with the fit quality found to be very good, consistent with~\cite{ATL-PHYS-PUB-2018-017}, and indeed in the $M_{tt}$ case in particular the fit quality is anomalously low, suggesting that the experimental errors may be overestimated. The impact on the gluon PDF is shown in Fig.~\ref{dilep}, and are seen to be broadly consistent within errors, though the rapidity prefers a somewhat lower gluon. In both cases some reduction in the PDF uncertainty at higher $x$ is seen, comparable in size with the same distributions in the lepton + jet channels. Indeed, it is interesting to note that the $y_{tt}$ distribution has a similar pull to the individual and combined decorrelated fits for the lepton + jet channel. 

We can therefore see that the issues present in the lepton + jet case are not apparent in the dilepton channel. Indeed, this may be natural in light of the fact that the impact of the type of two point MC uncertainties present in the lepton + jet case should be somewhat smaller in this cleaner channel. On the other hand, given that the most significant issues in the lepton + jet case came about when one considered a combined fit to all available distributions, it is difficult to make a firm statement without performing a combined fit to the $y_{tt}$ and $M_{tt}$ distributions. On the other hand, the fact that the fit quality is rather low in the case of the rapidity, and anomalously low in the case of the $M_{tt}$ distribution warrants some further investigation.

\section{Lepton + jets channel at CMS}\label{sec:CMS}

The final data set to be considered in this analysis is the lepton + jets channel measured by CMS~\cite{CMS_lep_jet}. While statistical correlations are provided within each distribution, they are not available across the different distributions, and hence fits can only be performed on these individually. Further to these correlations, we note that CMS provide a break-down for each systematic error in each bin into individual sources, which as in~\cite{NNPDF,Hou:2019gfw} we treat as a set of correlated errors across the individual distribution, one for each source.
However, with our findings from the ATLAS case in mind, we compare to the case that these systematic errors are instead treated as uncorrelated. We note in particular that the values for the errors given in the breakdown are all positive, this strongly calls into question their interpretation as a correlated source of errors across the bins of a normalized distributions.
The data are only presented as normalized distributions, and hence we multiply by the corresponding total cross section measurement~\cite{Khachatryan:2016mqs} to translate to the absolute case here. We treat the total systematic, statistical and luminosity uncertainty as additional sources of correlated systematic in all cases. Finally, as in~\cite{Sirunyan:2017azo} we remove the final bin in each distribution so that the covariance matrices corresponding to these normalised distributions are non--singular.

\begin{figure}
\centering
\begin{subfigure}{0.45\textwidth}
\includegraphics[width=\textwidth]{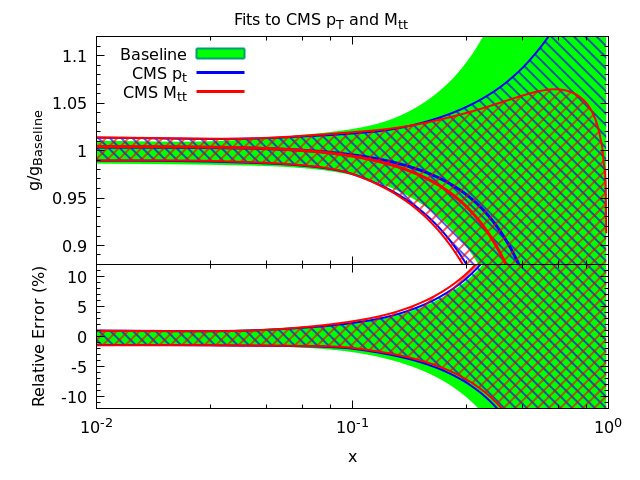}
\end{subfigure}
~
\begin{subfigure}{0.45\textwidth}
\includegraphics[width=\textwidth]{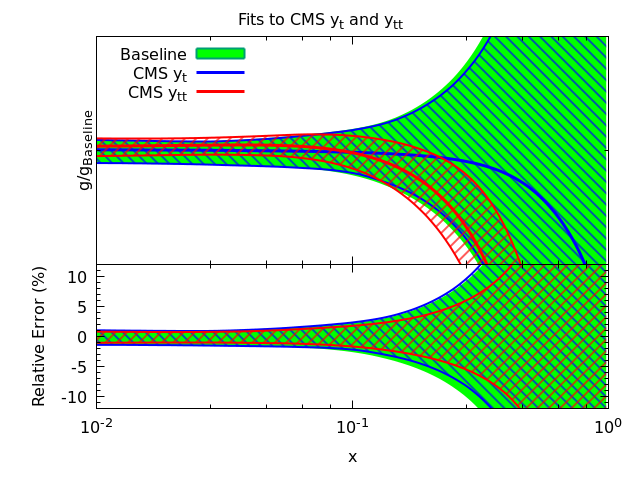}
\end{subfigure}
\caption{Impact of the CMS 8 TeV lepton+jets data on the gluon PDF.}
\label{cms}
\centering
\end{figure}

The results of these fits are shown in Table \ref{Table:cms}. The fit quality is fair in the case of the $y_{tt}$ distribution, while for the others it is noticeably worse, particularly for the $M_{tt}$. Interestingly, if we assume the systematic errors as uncorrelated we see a dramatic improvement in the fits to $p_\perp$, $y_t$ and $M_{tt}$, with some improvement in $y_{tt}.$ We leave a more detailed analysis of these effects for further study, in particular given the possible questions about the precise degree of correlation one should assume. The impact on the gluon PDF is shown in Fig.~\ref{cms}. We can see that all distributions have an impact at high $x$, with the $y_{tt}$ distribution having a larger constraining power. We also note that, in all cases, the pull is in the same direction as the ATLAS rapidity distributions, see Figs.~\ref{pdf 1} and \ref{dilep}.

\begin {table}
\begin{center}
\begin{tabular}{|c|c|c|} 
\hline
Distribution & Correlated & Uncorrelated \\
\hline
$p_{\perp}$ & 3.14 & 1.49 \\
\hline
$y_{t}$ & 2.71 & 1.25 \\
\hline
$y_{tt}$ & 1.70 & 1.39 \\
\hline
$M_{tt}$ & 5.81 & 2.97\\
\hline
\end{tabular}
\caption{$\chi^2/N_{\rm data}$ values for fits to the CMS 8 TeV lepton + jets data with the systematic errors taken as correlated and uncorrelated, as described in the text.}
\label{Table:cms} 
\end{center}
\end{table}

\section{Combining data sets}\label{sec:comb}
Finally we move on to combining the data sets already discussed into a final fit. We take the combination chosen by ATLAS in \cite{ATL-PHYS-PUB-2018-017}, namely the $M_{tt}$ and $p_\perp^t$ distributions, with the parton shower error decorrelated, from the lepton + jets channel and the $y_{tt}$ distribution from the dilepton channel (henceforth referred to as ATLAS default). On top of this, we add in the CMS distributions measured through the lepton + jets channel individually. The results for these are presented in Table \ref{Table:final}.

\begin{figure}
\centering
\begin{subfigure}{0.45\textwidth}
\includegraphics[width=\textwidth]{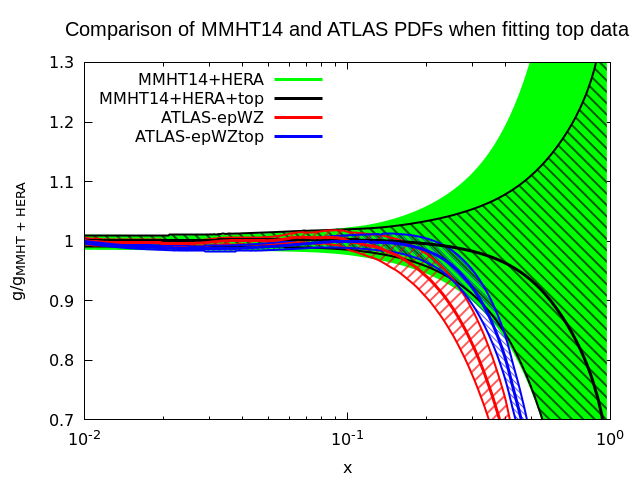}
\end{subfigure}
\caption{Impact of the ATLAS default combination on the MMHT14 and ATLAS-epWZ~\cite{Aaboud:2016btc} PDF sets.}
\label{atlas comp}
\centering
\end{figure}

The `ATLAS default' combination was chosen so as to be in line with the choice made by ATLAS in~\cite{ATL-PHYS-PUB-2018-017}, and as such a direct comparison of the corresponding PDF impact can be made as shown in Fig.~\ref{atlas comp}. Here we see that at high-$x$, the $t\bar{t}$ data causes the two baseline PDF sets (which themselves fit to rather different datasets) to converge. While they still disagree with each other in terms of error bands at rather high $x$, this is certainly encouraging to see.
\begin {table}
\begin{center}
\begin{tabular}{|c|c|c|c|c|} 
\hline
& $p_\perp^t + M_{tt}$ ATLAS & $y_{tt}$ ATLAS & CMS & \\
Distribution & lepton+jets (0.66) & dilepton (0.66) & lepton+jets & Total \\
\hline
ATLAS default & 0.74 & 1.26 & N/A & 0.87 \\
\hline
ATLAS default + CMS $p_{\perp}$ & 0.83 & 0.93 & 3.13 (3.14) & 1.45\\
\hline
ATLAS default + CMS $y_{t}$ & 0.74 & 1.27 & 2.71 (2.17) & 1.44\\
\hline
ATLAS default + CMS $y_{tt}$ & 0.90 & 0.63 & 1.79 (1.70) & 1.13 \\
\hline
ATLAS default + CMS $M_{tt}$ & 0.86 & 0.85 & 5.87 (5.81) & 2.02\\
\hline
\end{tabular}
\caption{$\chi^2/N_{\rm data}$ breakdown for combined fits to ATLAS and CMS $t\bar{t}$ data. The fourth column corresponds to the relevant rapidity distribution from CMS that is used in the fit. The numbers in brackets indicate the $\chi^2/N_{\rm data}$ when this distribution is fit individually.}
\label{Table:final} 
\end{center}
\end{table}
\begin{figure}
\centering
\begin{subfigure}{0.45\textwidth}
\includegraphics[width=\textwidth]{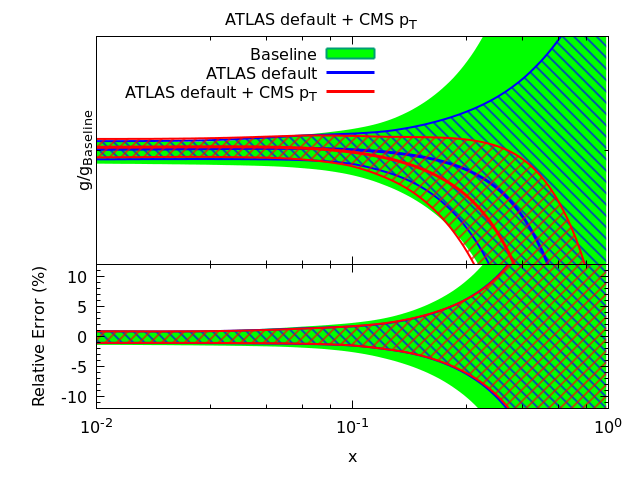}
\end{subfigure}
~
\begin{subfigure}{0.45\textwidth}
\includegraphics[width=\textwidth]{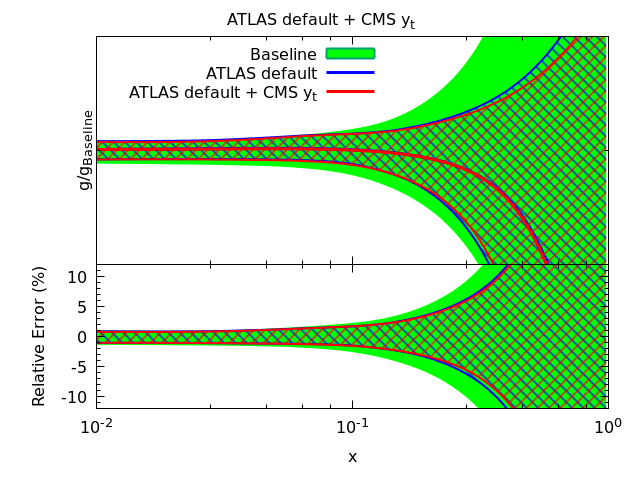}
\end{subfigure}
\begin{subfigure}{0.45\textwidth}
\includegraphics[width=\textwidth]{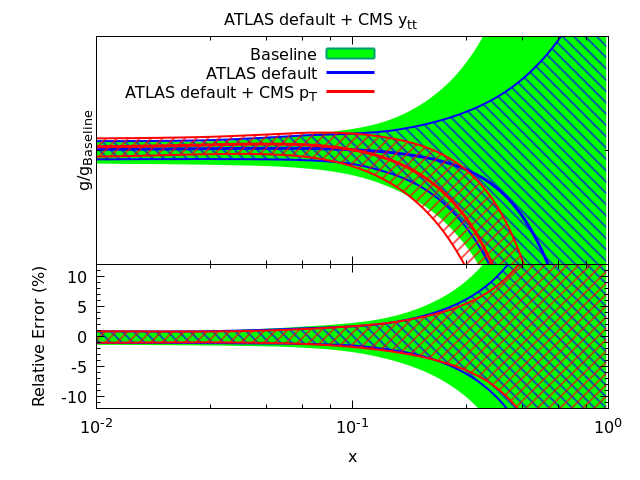}
\end{subfigure}
~
\begin{subfigure}{0.45\textwidth}
\includegraphics[width=\textwidth]{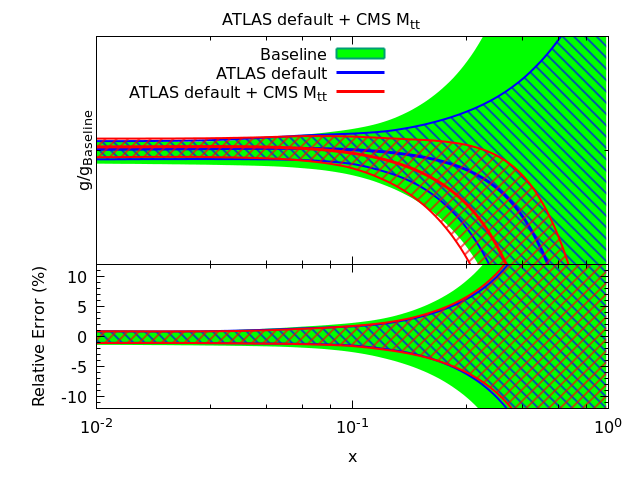}
\end{subfigure}
\caption{Impact of combining the default ATLAS combination of $p_T$ and $M_{tt}$ distributions with the parton shower error decorrelated from the lepton+jets channel and the $y_{tt}$ distribution from the dilepton with the CMS rapidity distributions on the gluon PDF.}
\label{atlas default}
\centering
\end{figure}

 The `ATLAS default' combination results in a deterioration in the ATLAS dilepton $y_{tt}$ distribution, however this is reduced upon adding in the CMS data in most cases, suggesting a good description of the ATLAS data can be achieved in conjunction with the CMS data. Interestingly this is not the case for the CMS $y_t$ distribution. 
 The impact of these final fits on the gluon is shown in Fig.~\ref{atlas default}, where we can see clearly that the ATLAS default combination provides good constraints on the high-$x$ gluon as expected. The CMS data further improves on the constraints imposed, with the $y_{tt}$ distribution having a stronger impact as before, while the others have very minimal constraining power. Considering the dynamically determined tolerances (see~\cite{Martin:2009iq} for a detailed description), we find that three eigenvector directions are constrained by the ATLAS dilepton and one by the ATLAS lepton + jets data in most of the fits to the ATLAS default combination, i.e. both with and without the CMS data. The only exception to this is the fit to the CMS $y_{tt}$ distribution, which confines three directions with the ATLAS dilepton confining another three, further demonstrating the constraining power of this distribution from CMS. Finally, we note that in all fits presented here, both individual and combined, no significant deterioration is observed in the fit quality to other data sets present in the fit, indicating that the $t\bar{t}$ data sets has no obvious tension with these. 

\section{Summary and Outlook}\label{sec:conc}

In this paper we have investigated in detail the effect of including LHC differential top quark production data within the global MMHT fit framework. Data for this have been provided by ATLAS and CMS in both the lepton + jet and dilepton decay channels, corrected back to the top quark parton level. These are then presented differentially in terms of various kinematic variables, namely the top quark pair invariant mass, $m_{t\overline{t}}$, and rapidity, $y_{t\overline{t}}$, and the individual top quark/antiquark transverse momentum, $p_\perp^t$, and rapidity, $y_t$. Such data are in particular a sensitive probe of the gluon PDF at high $x$.

The main focus of this paper has been on the ATLAS 8 TeV data presented in the lepton + jet channel, which is currently the only dataset for which the corresponding statistical and systematic error correlations have been provided across all distributions. The reason we concentrate on this is that it provides in principle the most constraining overall dataset, allowing all distributions to be fit at the same time without introducing the potential bias of choosing a particular distribution to fit. Moreover, this allows us to test whether full consistent dataset can in fact be described by the NNLO parton--level theory entering the PDF fit in a way that can only be partially done by considering individual distributions. 

We find, consistent with the ATLAS internal analysis, that only a very poor fit quality can be achieved to the full ATLAS dataset. This raises questions about the reliability of including such data in PDF fits, and in particular in simply choosing one distribution to fit, which would simply mask the underlying issues rather than resolving them. With this in mind we investigate the causes of this poor quality in detail, finding that it is dominated by the correlation prescription provided for the dominant sources of experimental error, namely two--point MC errors associated with the data unfolding back to parton--level; specifically due to the MC generator/input parameters used for the underlying hard scattering process, for the resulting parton shower and for ISR and FSR effects. We then discuss how the correlations provided for these errors across and within the different distributions may well be overly restrictive, and find that with some fairly mild loosening of these correlations, a significantly better fit quality can be achieved. Unfortunately, this has a non--negligible impact on the central value and uncertainties of the extracted high--$x$ gluon, resulting in particular in a rather smaller reduction in the corresponding PDF uncertainties after this decorrelation is introduced. Moreover, we explicitly compare this effect on the gluon with that of either excluding the NNLO QCD or NLO EW corrections to the top quark cross section, and find that it is significantly larger. Therefore, caution is required in claiming such data as it stands as a high precision probe of the gluon PDF.

We have in addition considered both the ATLAS 7 and 8 TeV data in the dilepton channel and the CMS 8 TeV data in the lepton + jet channel. In the former case we find a very good fit quality, with a reasonable impact on the gluon PDF from the top quark rapidity that is in fact rather similar to the pull from the corresponding distribution in the lepton + jet channel. This might indicate that the somewhat cleaner dilepton channel, for which the impact from the dominant systematic errors in the lepton + jet case should be smaller, may be a more promising channel to consider. However, it should be emphasised that as here the full experimental correlations across distributions have not been provided, there is a potential danger that similar underlying issues might be masked by the fits to individual distributions that we must necessarily perform. For the CMS data we have considered the 8 TeV lepton + jets channel, using the same distributions as in the ATLAS case. We find a reasonable description of the $y_{tt}$ distribution can be achieved, while the fit quality is poor for the others. These results are however highly sensitive to the treatment of the systematic error correlations. This is again found to be sensitive to the precise degree of correlation one assumes in the underlying systematic errors, though we have not studied this in detail here. Again, as the statistical correlations between the distributions are not available, we cannot perform a full analysis, including all distributions simultaneously.

In summary, we have found that the prescription for treating the correlated systematic errors in the case of the most comprehensive ATLAS lepton + jet data on differential top quark pair production plays a significant role in the overall constraining power of this dataset, which is in fact greater than the impact of either NNLO QCD or NLO EW corrections on the extracted gluon PDF. While we consider the above data explicitly, it should be emphasised that this effect is not limited to the ATLAS lepton + jet case, but may simply be masked in the other considered ATLAS and CMS datasets, where full correlations across distributions are not provided. We have introduced some reasonable procedures for decorrelating the dominant two--point MC uncertainties to judge the (rather large) impact on the fit, but this is clearly just a first step, rather than a firm prescription. Rather, a more complete understanding from both the experimental and theoretical point of view of the precision we can ascribe to the determined correlations of these uncertainties will be essential in the future. Moreover, the current common approach of fitting to an individual distribution risks introducing bias in the fit, and arguably will mask rather than resolve the above issues. Certainly, if the assumptions about the understanding of these correlations are too strong, we have found evidence that this may lead to an unreliable PDF determination, making a more conservative approach desirable. Alternatively, it might be that considering the relatively cleaner dilepton channel and/or alternative kinematic variables, perhaps say at the level of the decayed leptons rather than the corrected parton-level top quarks, might provide a cleaner probe of the gluon PDF. 

\section*{Acknowledgements}

We thank Amanda Cooper--Sarkar and Robert Thorne for useful discussions.
We thank Emanuele Nocera and C.-P. Yuan for providing information about the NNPDF and CT fits, and in particular the treatment of the systematic errors for the CMS data. S.~B. acknowledges financial support from the UK Science and Technology Facilities
Council. L.~H.~L thanks the Science and Technology Facilities Council (STFC)
for support via grant award ST/L000377/1.

\bibliography{ttbar_note_bib}{}
\bibliographystyle{h-physrev}

\end{document}